\documentclass[11pt]{article}
\usepackage[dvips]{graphicx}
\usepackage{amssymb}
\usepackage{color}
\usepackage{amsmath}
\usepackage{nicefrac}
\usepackage[left]{lineno}
\usepackage{hyperref}
\usepackage{adjustbox}
\usepackage{xspace}
\usepackage{multirow}

\definecolor{burgundy}{rgb}{0.5, 0.0, 0.13}

\def\geant {\mbox{\textsc{Geant4}}\xspace}

\begin{document}
\centerline{\LARGE EUROPEAN ORGANIZATION FOR NUCLEAR RESEARCH}
%
\vspace{10mm} {\flushright{
CERN-EP-2020-089 \\
19 May 2020\\
\vspace{4mm}
Revised version:\\19 June 2020\\
}}
\vspace{-30mm}

%
%

%
\vspace{40mm}

\begin{center}
\boldmath
{\bf {\Large\boldmath{Search for heavy neutral lepton production\\in $K^+$ decays to positrons}}}
\unboldmath
\end{center}
\begin{center}
{\Large The NA62 Collaboration}\\
\end{center}

\begin{abstract}
A search for heavy neutral lepton ($N$) production in $K^+\to e^+N$ decays using the data sample collected by the NA62 experiment at CERN in 2017--2018 is reported. Upper limits of the extended neutrino mixing matrix element $|U_{e4}|^2$ are established at the level of $10^{-9}$ over most of the accessible heavy neutral lepton mass range 144--462~MeV/$c^2$, with the assumption that the lifetime exceeds 50~ns. These limits improve significantly upon those of previous production and decay searches. The $|U_{e4}|^2$ range favoured by Big Bang Nucleosynthesis is excluded up to a mass of about 340~MeV/$c^2$.
\end{abstract}

\begin{center}
{\it Accepted for publication in Physics Letters B}
\end{center}

\newpage
 \begin{center}
{\Large The NA62 Collaboration$\,$\renewcommand{\thefootnote}{\fnsymbol{footnote}}%
\footnotemark[1]\renewcommand{\thefootnote}{\arabic{footnote}}}\\
\end{center}
 \vspace{3mm}
\begin{raggedright}
\noindent

{\bf Universit\'e Catholique de Louvain, Louvain-La-Neuve, Belgium}\\
 E.~Cortina Gil,
 A.~Kleimenova,
 E.~Minucci$\,$\footnotemark[1]$^,\,$\footnotemark[2],
 S.~Padolski$\,$\footnotemark[3],
 P.~Petrov,
 A.~Shaikhiev$\,$\footnotemark[4],
 R.~Volpe$\,$\footnotemark[5]\\[2mm]

{\bf TRIUMF, Vancouver, British Columbia, Canada}\\
 T.~Numao,
 Y.~Petrov,
 B.~Velghe\\[2mm]

{\bf University of British Columbia, Vancouver, British Columbia, Canada}\\
 D.~Bryman$\,$\footnotemark[6],
 J.~Fu$\,$\footnotemark[7]\\[2mm]

{\bf Charles University, Prague, Czech Republic}\\
 T.~Husek$\,$\footnotemark[8],
 J.~Jerhot$\,$\footnotemark[9],
 K.~Kampf,
 M.~Zamkovsky\\[2mm]

{\bf Institut f\"ur Physik and PRISMA Cluster of excellence, Universit\"at Mainz, Mainz, Germany}\\
 R.~Aliberti$\,$\footnotemark[10],
 G.~Khoriauli$\,$\footnotemark[11],
 J.~Kunze,
 D.~Lomidze$\,$\footnotemark[12],
 L.~Peruzzo,
 M.~Vormstein,
 R.~Wanke\\[2mm]

{\bf Dipartimento di Fisica e Scienze della Terra dell'Universit\`a e INFN, Sezione di Ferrara, Ferrara, Italy}\\
 P.~Dalpiaz,
 M.~Fiorini,
 I.~Neri,
 A.~Norton,
 F.~Petrucci,
 H.~Wahl\\[2mm]

{\bf INFN, Sezione di Ferrara, Ferrara, Italy}\\
 A.~Cotta Ramusino,
 A.~Gianoli\\[2mm]

{\bf Dipartimento di Fisica e Astronomia dell'Universit\`a e INFN, Sezione di Firenze, Sesto Fiorentino, Italy}\\
 E.~Iacopini,
 G.~Latino,
 M.~Lenti,
 A.~Parenti\\[2mm]

{\bf INFN, Sezione di Firenze, Sesto Fiorentino, Italy}\\
 A.~Bizzeti$\,$\footnotemark[13],
 F.~Bucci\\[2mm]

{\bf Laboratori Nazionali di Frascati, Frascati, Italy}\\
 A.~Antonelli,
 G.~Georgiev$\,$\footnotemark[14],
 V.~Kozhuharov$\,$\footnotemark[14],
 G.~Lanfranchi,
 S.~Martellotti,
 M.~Moulson,
 T.~Spadaro\\[2mm]

{\bf Dipartimento di Fisica ``Ettore Pancini'' e INFN, Sezione di Napoli, Napoli, Italy}\\
 F.~Ambrosino,
 T.~Capussela,
 M.~Corvino,
 D.~Di Filippo,
 P.~Massarotti,
 M.~Mirra,
 M.~Napolitano,
 G.~Saracino\\[2mm]

{\bf Dipartimento di Fisica e Geologia dell'Universit\`a e INFN, Sezione di Perugia, Perugia, Italy}\\
 G.~Anzivino,
 F.~Brizioli,
 E.~Imbergamo,
 R.~Lollini,
 R.~Piandani,
 C.~Santoni\\[2mm]

{\bf INFN, Sezione di Perugia, Perugia, Italy}\\
 M.~Barbanera$\,$\footnotemark[15],
 P.~Cenci,
 B.~Checcucci,
 P.~Lubrano,
 M.~Lupi$\,$\footnotemark[16],
 M.~Pepe,
 M.~Piccini\\[2mm]

{\bf Dipartimento di Fisica dell'Universit\`a e INFN, Sezione di Pisa, Pisa, Italy}\\
 F.~Costantini,
 L.~Di Lella,
 N.~Doble,
 M.~Giorgi,
 S.~Giudici,
 G.~Lamanna,
 E.~Lari,
 E.~Pedreschi,
 M.~Sozzi\\[2mm]

{\bf INFN, Sezione di Pisa, Pisa, Italy}\\
 C.~Cerri,
 R.~Fantechi,
 L.~Pontisso,
 F.~Spinella\\[2mm]

{\bf Scuola Normale Superiore e INFN, Sezione di Pisa, Pisa, Italy}\\
 I.~Mannelli\\[2mm]

{\bf Dipartimento di Fisica, Sapienza Universit\`a di Roma e INFN, Sezione di Roma I, Roma, Italy}\\
 G.~D'Agostini,
 M.~Raggi\\[2mm]

{\bf INFN, Sezione di Roma I, Roma, Italy}\\
 A.~Biagioni,
 E.~Leonardi,
 A.~Lonardo,
 P.~Valente,
 P.~Vicini\\[2mm]

{\bf INFN, Sezione di Roma Tor Vergata, Roma, Italy}\\
 R.~Ammendola,
 V.~Bonaiuto$\,$\footnotemark[17],
 A.~Fucci,
 A.~Salamon,
 F.~Sargeni$\,$\footnotemark[18]\\[2mm]

{\bf Dipartimento di Fisica dell'Universit\`a e INFN, Sezione di Torino, Torino, Italy}\\
 R.~Arcidiacono$\,$\footnotemark[19],
 B.~Bloch-Devaux,
 M.~Boretto$\,$\footnotemark[20],
 E.~Menichetti,
 E.~Migliore,
 D.~Soldi\\[2mm]

{\bf INFN, Sezione di Torino, Torino, Italy}\\
 C.~Biino,
 A.~Filippi,
 F.~Marchetto\\[2mm]

{\bf Instituto de F\'isica, Universidad Aut\'onoma de San Luis Potos\'i, San Luis Potos\'i, Mexico}\\
 J.~Engelfried,
 N.~Estrada-Tristan$\,$\footnotemark[21]\\[2mm]

{\bf Horia Hulubei National Institute of Physics for R\&D in Physics and Nuclear Engineering, Bucharest-Magurele, Romania}\\
 A. M.~Bragadireanu,
 S. A.~Ghinescu,
 O. E.~Hutanu\\[2mm]

{\bf Joint Institute for Nuclear Research, Dubna, Russia}\\
 A.~Baeva,
 D.~Baigarashev,
 D.~Emelyanov,
 T.~Enik,
 V.~Falaleev,
 V.~Kekelidze,
 A.~Korotkova,
 L.~Litov$\,$\footnotemark[14],
 D.~Madigozhin,
 M.~Misheva$\,$\footnotemark[22],
 N.~Molokanova,
 S.~Movchan,
 I.~Polenkevich,
 Yu.~Potrebenikov,
 S.~Shkarovskiy,
 A.~Zinchenko$\,$\renewcommand{\thefootnote}{\fnsymbol{footnote}}\footnotemark[2]\renewcommand{\thefootnote}{\arabic{footnote}}\\[2mm]

{\bf Institute for Nuclear Research of the Russian Academy of Sciences, Moscow, Russia}\\
 S.~Fedotov,
 E.~Gushchin,
 A.~Khotyantsev,
 Y.~Kudenko$\,$\footnotemark[23],
 V.~Kurochka,
 M.~Medvedeva,
 A.~Mefodev\\[2mm]

{\bf Institute for High Energy Physics - State Research Center of Russian Federation, Protvino, Russia}\\
 S.~Kholodenko,
 V.~Kurshetsov,
 V.~Obraztsov,
 A.~Ostankov$\,$\renewcommand{\thefootnote}{\fnsymbol{footnote}}\footnotemark[2]\renewcommand{\thefootnote}{\arabic{footnote}},
 V.~Semenov$\,$\renewcommand{\thefootnote}{\fnsymbol{footnote}}\footnotemark[2]\renewcommand{\thefootnote}{\arabic{footnote}},
 V.~Sugonyaev,
 O.~Yushchenko\\[2mm]

{\bf Faculty of Mathematics, Physics and Informatics, Comenius University, Bratislava, Slovakia}\\
 L.~Bician$\,$\footnotemark[20],
 T.~Blazek,
 V.~Cerny,
 Z.~Kucerova\\[2mm]

{\bf CERN,  European Organization for Nuclear Research, Geneva, Switzerland}\\
 J.~Bernhard,
 A.~Ceccucci,
 H.~Danielsson,
 N.~De Simone$\,$\footnotemark[24],
 F.~Duval,
 B.~D\"obrich,
 L.~Federici,
 E.~Gamberini,
 L.~Gatignon,
 R.~Guida,
 F.~Hahn$\,$\renewcommand{\thefootnote}{\fnsymbol{footnote}}\footnotemark[2]\renewcommand{\thefootnote}{\arabic{footnote}},
 E. B.~Holzer,
 B.~Jenninger,
 M.~Koval$\,$\footnotemark[25],
 P.~Laycock$\,$\footnotemark[3],
 G.~Lehmann Miotto,
 P.~Lichard,
 A.~Mapelli,
 R.~Marchevski,
 K.~Massri,
 M.~Noy,
 V.~Palladino$\,$\footnotemark[26],
 M.~Perrin-Terrin$\,$\footnotemark[27]$^,\,$\footnotemark[28],
 J.~Pinzino$\,$\footnotemark[29]$^,\,$\footnotemark[30],
 V.~Ryjov,
 S.~Schuchmann$\,$\footnotemark[31],
 S.~Venditti\\[2mm]

{\bf University of Birmingham, Birmingham, United Kingdom}\\
 T.~Bache,
 M. B.~Brunetti$\,$\footnotemark[32],
 V.~Duk$\,$\footnotemark[33],
 V.~Fascianelli$\,$\footnotemark[34],
 J. R.~Fry,
 F.~Gonnella,
 E.~Goudzovski$\,$\renewcommand{\thefootnote}{\fnsymbol{footnote}}%
\footnotemark[1]\renewcommand{\thefootnote}{\arabic{footnote}},
 J.~Henshaw,
 L.~Iacobuzio,
 C.~Lazzeroni,
 N.~Lurkin$\,$\footnotemark[28],
 F.~Newson,
 C.~Parkinson$\,$\footnotemark[9],
 A.~Romano,
 A.~Sergi,
 A.~Sturgess,
 J.~Swallow\\[2mm]

{\bf University of Bristol, Bristol, United Kingdom}\\
 H.~Heath,
 R.~Page,
 S.~Trilov\\[2mm]
\newpage
{\bf University of Glasgow, Glasgow, United Kingdom}\\
 B.~Angelucci,
 D.~Britton,
 C.~Graham,
 D.~Protopopescu\\[2mm]

{\bf University of Lancaster, Lancaster, United Kingdom}\\
 J.~Carmignani,
 J. B.~Dainton,
 R. W. L.~Jones,
 G.~Ruggiero\\[2mm]

{\bf University of Liverpool, Liverpool, United Kingdom}\\
 L.~Fulton,
 D.~Hutchcroft,
 E.~Maurice$\,$\footnotemark[35],
 B.~Wrona\\[2mm]

{\bf George Mason University, Fairfax, Virginia, USA}\\
 A.~Conovaloff,
 P.~Cooper,
 D.~Coward$\,$\footnotemark[36],
 P.~Rubin\\[2mm]

\end{raggedright}
%
%
\setcounter{footnote}{0}
\renewcommand{\thefootnote}{\fnsymbol{footnote}}
\footnotetext[1]{Corresponding author: Evgueni Goudzovski,
email: Evgueni.Goudzovski@cern.ch}
\footnotetext[2]{Deceased}
\renewcommand{\thefootnote}{\arabic{footnote}}

\footnotetext[1]{Present address: Laboratori Nazionali di Frascati, I-00044 Frascati, Italy}
\footnotetext[2]{Also at CERN,  European Organization for Nuclear Research, CH-1211 Geneva 23, Switzerland}
\footnotetext[3]{Present address: Brookhaven National Laboratory, Upton, NY 11973, USA}
\footnotetext[4]{Also at Institute for Nuclear Research of the Russian Academy of Sciences, 117312 Moscow, Russia}
\footnotetext[5]{Present address: Faculty of Mathematics, Physics and Informatics, Comenius University, 842 48, Bratislava, Slovakia}
\footnotetext[6]{Also at TRIUMF, Vancouver, British Columbia, V6T 2A3, Canada}
\footnotetext[7]{Present address: UCLA Physics and Biology in Medicine, Los Angeles, CA 90095, USA}
\footnotetext[8]{Present address: IFIC, Universitat de Val\`encia - CSIC, E-46071 Val\`encia, Spain}
\footnotetext[9]{Present address: Universit\'e Catholique de Louvain, B-1348 Louvain-La-Neuve, Belgium}
\footnotetext[10]{Present address: Institut f\"ur Kernphysik and Helmholtz Institute Mainz, Universit\"at Mainz, Mainz, D-55099, Germany}
\footnotetext[11]{Present address: Universit\"at W\"urzburg, D-97070 W\"urzburg, Germany}
\footnotetext[12]{Present address: Universit\"at Hamburg, D-20146 Hamburg, Germany}
\footnotetext[13]{Also at Dipartimento di Fisica, Universit\`a di Modena e Reggio Emilia, I-41125 Modena, Italy}
\footnotetext[14]{Also at Faculty of Physics, University of Sofia, BG-1164 Sofia, Bulgaria}
\footnotetext[15]{Present address: INFN, Sezione di Pisa, I-56100 Pisa, Italy}
\footnotetext[16]{Present address: Institut am Fachbereich Informatik und Mathematik, Goethe Universit\"at, D-60323 Frankfurt am Main, Germany}
\footnotetext[17]{Also at Department of Industrial Engineering, University of Roma Tor Vergata, I-00173 Roma, Italy}
\footnotetext[18]{Also at Department of Electronic Engineering, University of Roma Tor Vergata, I-00173 Roma, Italy}
\footnotetext[19]{Also at Universit\`a degli Studi del Piemonte Orientale, I-13100 Vercelli, Italy}
\footnotetext[20]{Present address: CERN,  European Organization for Nuclear Research, CH-1211 Geneva 23, Switzerland}
\footnotetext[21]{Also at Universidad de Guanajuato, Guanajuato, Mexico}
\footnotetext[22]{Present address: Institute of Nuclear Research and Nuclear Energy of Bulgarian Academy of Science (INRNE-BAS), BG-1784 Sofia, Bulgaria}
\footnotetext[23]{Also at National Research Nuclear University (MEPhI), 115409 Moscow and Moscow Institute of Physics and Technology, 141701 Moscow region, Moscow, Russia}
\footnotetext[24]{Present address: DESY, D-15738 Zeuthen, Germany}
\footnotetext[25]{Present address: Charles University, 116 36 Prague 1, Czech Republic}
\footnotetext[26]{Present address: Physics Department, Imperial College London, London, SW7 2BW, UK}
\footnotetext[27]{Present address: Centre de Physique des Particules de Marseille, Universit\'e Aix Marseille, CNRS/IN2P3, F-13288, Marseille, France}
\footnotetext[28]{Also at Universit\'e Catholique de Louvain, B-1348 Louvain-La-Neuve, Belgium}
\footnotetext[29]{Present address: Department of Physics, University of Toronto, Toronto, Ontario, M5S 1A7, Canada}
\footnotetext[30]{Also at INFN, Sezione di Pisa, I-56100 Pisa, Italy}
\footnotetext[31]{Present address: Institut f\"ur Physik and PRISMA Cluster of excellence, Universit\"at Mainz, D-55099 Mainz, Germany}
\footnotetext[32]{Present address: Department of Physics, University of Warwick, Coventry, CV4 7AL, UK}
\footnotetext[33]{Present address: INFN, Sezione di Perugia, I-06100 Perugia, Italy}
\footnotetext[34]{Present address: Dipartimento di Psicologia, Universit\`a di Roma La Sapienza, I-00185 Roma, Italy}
\footnotetext[35]{Present address: Laboratoire Leprince Ringuet, F-91120 Palaiseau, France}
\footnotetext[36]{Also at SLAC National Accelerator Laboratory, Stanford University, Menlo Park, CA 94025, USA}

\newpage


\section*{Introduction}

All Standard Model (SM) fermions except neutrinos are known to exhibit right-handed chirality. The existence of right-handed neutrinos, or heavy neutral leptons (HNLs), is hypothesised in many SM extensions in order to generate non-zero masses of the SM neutrinos via the seesaw mechanism~\cite{pbc19}. For example, the Neutrino Minimal Standard Model~\cite{nuMSM} simultaneously accounts for dark matter, baryogenesis, and neutrino masses and oscillations by postulating two HNLs in the MeV--GeV mass range and a third HNL, a dark matter candidate, at the keV mass scale.

Mixing between HNLs (also denoted $N$ below) and active neutrinos gives rise to HNL production in decays of SM particles and decays of HNLs into SM particles. Both classes of processes can in principle be detected experimentally. The expected branching fraction of the $K^+\to\ell^+N$ decay ($\ell=e,\mu$) is~\cite{sh80}
\begin{equation}
{\cal B}(K^+\to\ell^+ N) = {\cal B}(K^+\to\ell^+\nu) \cdot \rho_\ell(m_N) \cdot |U_{\ell 4}|^2,
\label{eq:main}
\end{equation}
where ${\cal B}(K^+\to\ell^+\nu)$ is the measured branching fraction of the SM leptonic decay, $|U_{\ell 4}|^2$ is the mixing parameter, $m_N$ is the HNL mass, and $\rho_\ell(m_N)$ is a kinematic factor:
\begin{equation}
\label{eq:rho}
\rho_\ell(m_N) = \frac {(x+y)-(x-y)^2} {x(1-x)^2} \cdot \lambda^{1/2}(1,x,y),
\end{equation}
with $x=(m_\ell/m_K)^2$, $y=(m_N/m_K)^2$ and $\lambda(a,b,c)=a^2+b^2+c^2-2(ab+bc+ac)$. By definition, $\rho_\ell(0)=1$. Numerically, the product ${\cal B}(K^+\to\ell^+\nu) \cdot \rho_\ell(m_N)$ is ${\cal O}(1)$ over most of the allowed $m_N$ range; it drops to zero at the kinematic limit $m_N=m_K-m_\ell$ and, in the positron case, reduces to ${\cal B}(K^+\to e^+\nu)=1.582(7)\times 10^{-5}$~\cite{pdg} for $m_N \to 0$ due to helicity suppression.


The lifetime of an HNL with mass $m_N < m_K$ and decaying exclusively into SM particles will exceed $10^{-4}/|U_4|^2~\mu$s, where $|U_4|^2$ is the largest of the three coupling parameters $U_{\ell 4}^2$ ($\ell=e,\mu,\tau$)~\cite{bo18}. Assuming conservatively that $|U_{\ell 4}|^2<10^{-4}$, the lifetime exceeds 1~$\mu$s and HNLs can be considered stable in production-search experiments.

A search for $K^+\to e^+N$ decays in the HNL mass range 144--462~MeV/$c^2$ using the data collected by the NA62 experiment at CERN in 2017--2018 is reported. The results, which assume the HNL lifetime exceeds 50~ns, are presented as upper limits of $|U_{e4}|^2$ at 90\% CL for a number of mass hypotheses.

\section{Beam, detector and data sample}
\label{sec:detector}

\begin{figure}[t]
\begin{center}
\resizebox{\textwidth}{!}{\includegraphics{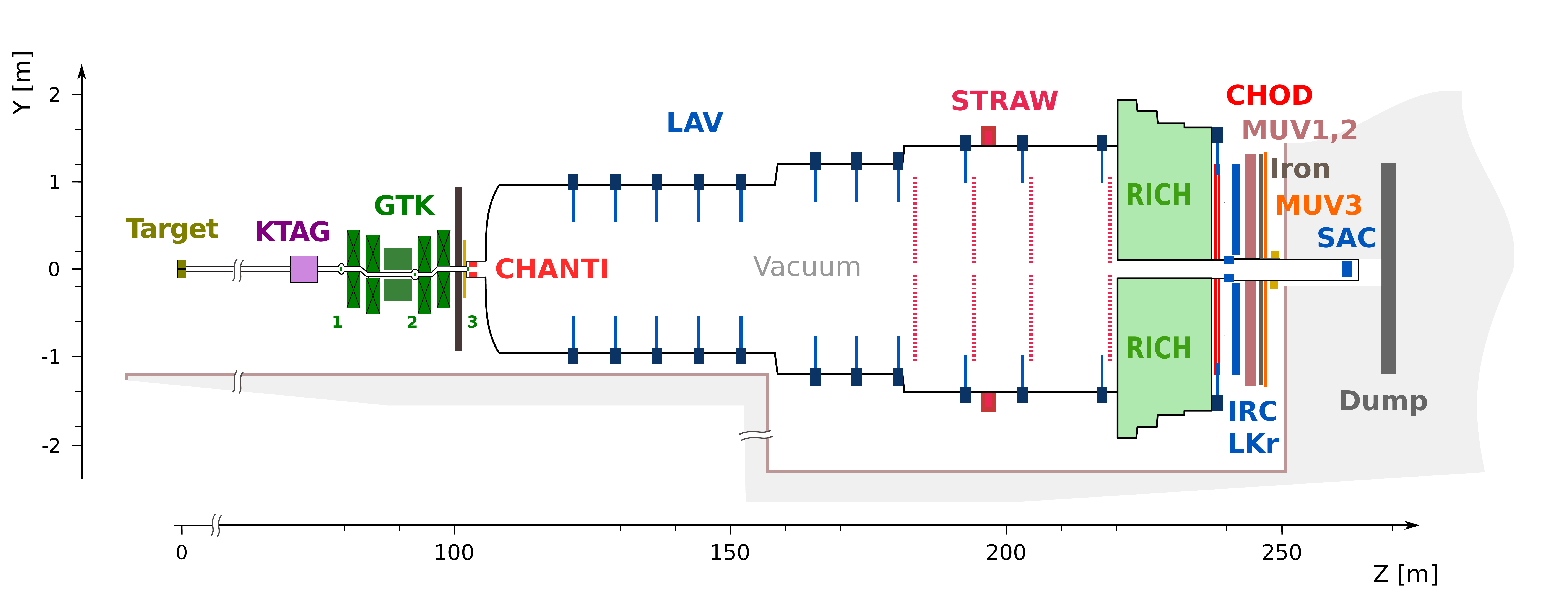}}
\put(-342,67){\tiny\color{burgundy}\rotatebox{90}{\bf SCR}}
\put(-324,121){\tiny\color{burgundy}\rotatebox{90}{\bf COL}}
\put(-172,46){\scriptsize\color{burgundy}{\bf M}}
\end{center}
\vspace{-16mm}
\caption{Schematic side view of the NA62 beamline and detector.}
\label{fig:detector}
\end{figure}

The layout of the NA62 beamline and detector~\cite{na62-detector} is shown schematically in Fig.~\ref{fig:detector}. An unseparated secondary beam of $\pi^+$ (70\%), protons (23\%) and $K^+$ (6\%) is created by directing 400~GeV/$c$ protons extracted from the CERN SPS onto a beryllium target in spills of 3~s effective duration. The central beam momentum is 75~GeV/$c$, with a momentum spread of 1\% (rms).

Beam kaons are tagged with 70~ps time resolution by a differential Cherenkov counter (KTAG) using nitrogen gas at 1.75~bar pressure contained in a 5~m long vessel as radiator. Beam particle positions, momenta and times (to better than 100~ps resolution) are measured by a silicon pixel spectrometer consisting of three stations (GTK1,2,3) and four dipole magnets. A muon scraper (SCR) is installed between GTK1 and GTK2. A 1.2~m thick steel collimator (COL) with a central aperture of $76\times40$~mm$^2$ and outer dimensions of $1.7\times1.8$~m$^2$ is placed upstream of GTK3 to absorb hadrons from upstream $K^+$ decays (a variable aperture collimator of $0.15\times0.15$~m$^2$ outer dimensions was used up to early 2018). Inelastic interactions of beam particles in GTK3 are detected by an array of scintillator hodoscopes (CHANTI) located just after GTK3. The beam is delivered into a vacuum tank evacuated to a pressure of $10^{-6}$~mbar, which contains a 75~m long fiducial decay volume (FV) starting 2.6~m downstream of GTK3. The beam divergence at the FV entrance is 0.11~mrad (rms) in both horizontal and vertical planes. Downstream of the FV, undecayed beam particles continue their path in vacuum.

Momenta of charged particles produced by $K^+$ decays in the FV are measured by a magnetic spectrometer (STRAW) located in the vacuum tank downstream of the FV. The spectrometer consists of four tracking chambers made of straw tubes, and a dipole magnet (M) located between the second and third chambers and providing a horizontal momentum kick of 270~MeV/$c$. The momentum resolution achieved is $\sigma_p/p = (0.30\oplus 0.005\cdot p)\%$, where the momentum $p$ is expressed in GeV/$c$.

A ring-imaging Cherenkov detector (RICH), consisting of a 17.5~m long vessel filled with neon at atmospheric pressure (with a Cherenkov threshold for muons of 9.5~GeV/$c$), is used for the identification of charged particles and time measurements with 70~ps precision (for positrons). Two scintillator hodoscopes (CHOD, which include a matrix of tiles and two orthogonal planes of slabs, arranged in four quadrants) downstream of the RICH provide trigger signals and time measurements with 200~ps precision.

A $27X_0$ thick quasi-homogeneous liquid krypton (LKr) electromagnetic calorimeter is used for particle identification and photon detection. The calorimeter has an active volume of 7~m$^3$, is segmented in the transverse direction into 13248 projective cells of approximately $2\!\times\!2$~cm$^2$, and provides an energy resolution $\sigma_E/E=(4.8/\sqrt{E}\oplus11/E\oplus0.9)\%$, where $E$ is expressed in GeV. To achieve hermetic acceptance for photons emitted in $K^+$ decays in the FV at angles up to 50~mrad to the beam axis, the LKr calorimeter is supplemented by annular lead glass detectors (LAV) installed in 12~positions in and downstream of the vacuum tank, and two lead/scintillator sampling calorimeters (IRC, SAC) located close to the beam axis. An iron/scintillator sampling hadronic calorimeter formed of two modules (MUV1,2) and a muon detector (MUV3) consisting of 148~scintillator tiles located behind an 80~cm thick iron wall are used for particle identification.

%
%

The data sample used for the analysis is obtained from $0.79\times 10^6$ SPS spills recorded during 360 days of operation in 2017--2018, at a typical beam intensity of \mbox{$2.2\times 10^{12}$} protons per spill corresponding to a mean beam particle rate at the FV entrance of 500~MHz, and a mean $K^+$ decay rate in the FV of 3.7~MHz. The trigger used for the $K^+\to\pi\nu\bar\nu$ measurement~\cite{co20}, consisting of both hardware (L0) and software (L1) stages, is used for the analysis. Overall trigger efficiency for single positrons with momenta below 30~GeV/$c$ is measured to be $(90\pm4)\%$ using data samples.


\section{Event selection}
\label{sec:selection}

Assuming an HNL lifetime exceeding 50~ns, and considering that the HNL produced in $K^+\to e^+N$ decays would be boosted by a Lorentz factor of ${\cal O}(100)$, HNL decays in flight into SM particles in the 156~m long volume between the start of the FV and the last detector (SAC) can be neglected (Section~\ref{sec:search}). Therefore the $K^+\to e^+N$ decay is characterized by a single positron in the final state, similarly to the SM $K^+\to e^+\nu$ decay. The principal selection criteria follow.
\begin{itemize}
\item A positron track reconstructed in the STRAW spectrometer with momentum in the range 5--30~GeV/$c$ is required. The momentum is restricted to this range, because the L0 trigger required that the total energy deposited in the LKr not exceed 30~GeV. The track's trajectory through the STRAW chambers and its extrapolation to the LKr calorimeter, CHOD and MUV3 should be within the fiducial geometrical acceptance of these detectors. The positron time is evaluated as the mean time of the RICH signals spatially associated with the track.
\item Backgrounds due to particle misidentification are suppressed to a negligible level by applying the following particle identification criteria to the single track: the ratio of energy, $E$, deposited in the LKr calorimeter to momentum, $p$, measured by the STRAW spectrometer is required to be \mbox{$0.92<E/p<1.08$}; a particle identification algorithm based on the RICH signal pattern within 3~ns of the positron RICH time is applied; no signal in the MUV3 detector spatially consistent with the projected track impact point and within 4~ns of the positron time is allowed.
\item Backgrounds from beam pion decays (mainly $\pi^+\to e^+\nu$, and to a lesser extent $\pi^+\to\mu^+\nu$ followed by muon decay $\mu^+\to e^+\nu\bar\nu$) are suppressed by requiring a kaon signal in the KTAG detector within 1~ns of the positron time.
\item Identification of the $K^+$ track in the GTK relies on the time difference, $\Delta t_{\rm GK}$, between a GTK track and the KTAG signal, and spatial compatibility of the GTK and STRAW tracks quantified by the distance, $d$, of closest approach. A discriminant ${\cal D}(\Delta t_{\rm GK}, d)$ is defined using the $\Delta t_{\rm GK}$ and $d$ distributions measured with $K^+\to\pi^+\pi^+\pi^-$ decays. Among GTK tracks with $|\Delta t_{\rm GK}|<0.5$~ns, the track of the parent kaon is assumed to be the one giving the ${\cal D}$ value most consistent with a $K^+\to e^+$ decay. It is also required that $d<4$~mm to reduce the background from $K^+\to\mu^+\nu$ decays in the FV followed by muon decay $\mu^+\to e^+\nu\bar\nu$. The decay vertex is defined as the point of closest approach of the GTK and STRAW tracks, taking into account the stray magnetic field in the vacuum tank.
\item Background from $K^+\to\mu^+\nu$ decays upstream of GTK3 followed by $\mu^+$ decays in the FV arises from pileup in the GTK, and is suppressed by geometrical conditions. Namely, the reconstructed $K^+$ decay vertex is required to be located in the FV at a minimum distance from the start of the FV, varying from 2~m to 12~m depending on the angle between the $K^+$ momentum in the laboratory frame and the positron momentum in the $K^+$ rest frame.
\item Backgrounds from multi-body $K^+$ decays are suppressed by veto conditions. The positron track must not form vertices with any other STRAW track. LKr energy deposition clusters not spatially compatible with the positron track within 8~ns of the positron time are not allowed. Activity in the large-angle (LAV) and small-angle (SAC, IRC) photon veto detectors within 3~ns of the positron time is forbidden. Activity in the CHANTI detector within 4~ns of the positron time is not allowed; more than two signals in the CHOD tiles within 6~ns of the positron time and not spatially associated with the positron are also not allowed. Data loss due to the veto conditions from accidental activity (``random veto'') averaged over the sample is measured to be about 30\%.
\end{itemize}

The squared missing mass is computed as $m_{\rm miss}^2=(P_K-P_e)^2$, where $P_K$ and $P_e$ are the kaon and positron 4-momenta, obtained from the 3-momenta measured by the GTK and STRAW detectors and using the $K^+$ and $e^+$ mass hypotheses.

\begin{figure}[t]
\begin{center}
\resizebox{0.50\textwidth}{!}{\includegraphics{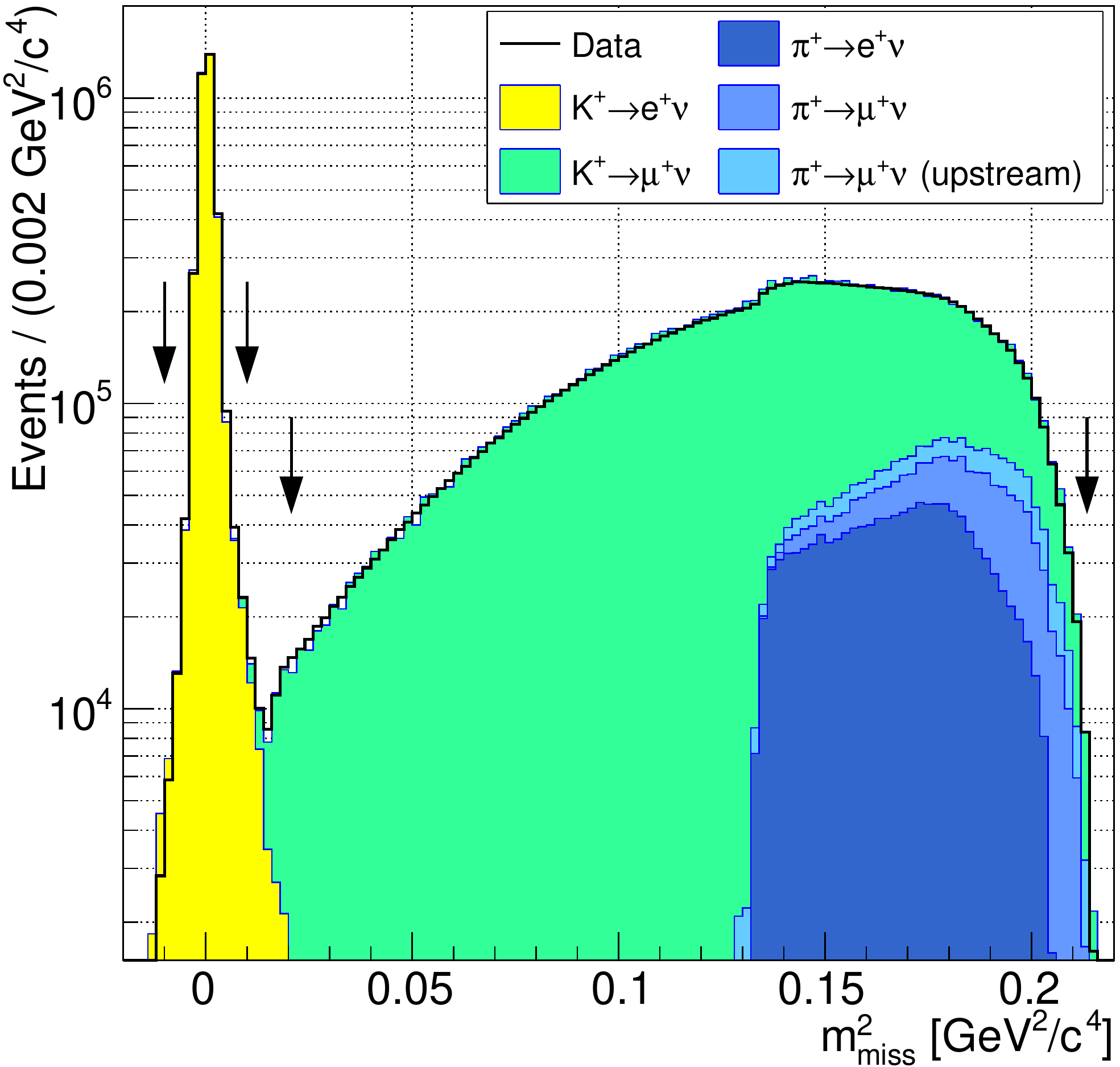}}%
\resizebox{0.50\textwidth}{!}{\includegraphics{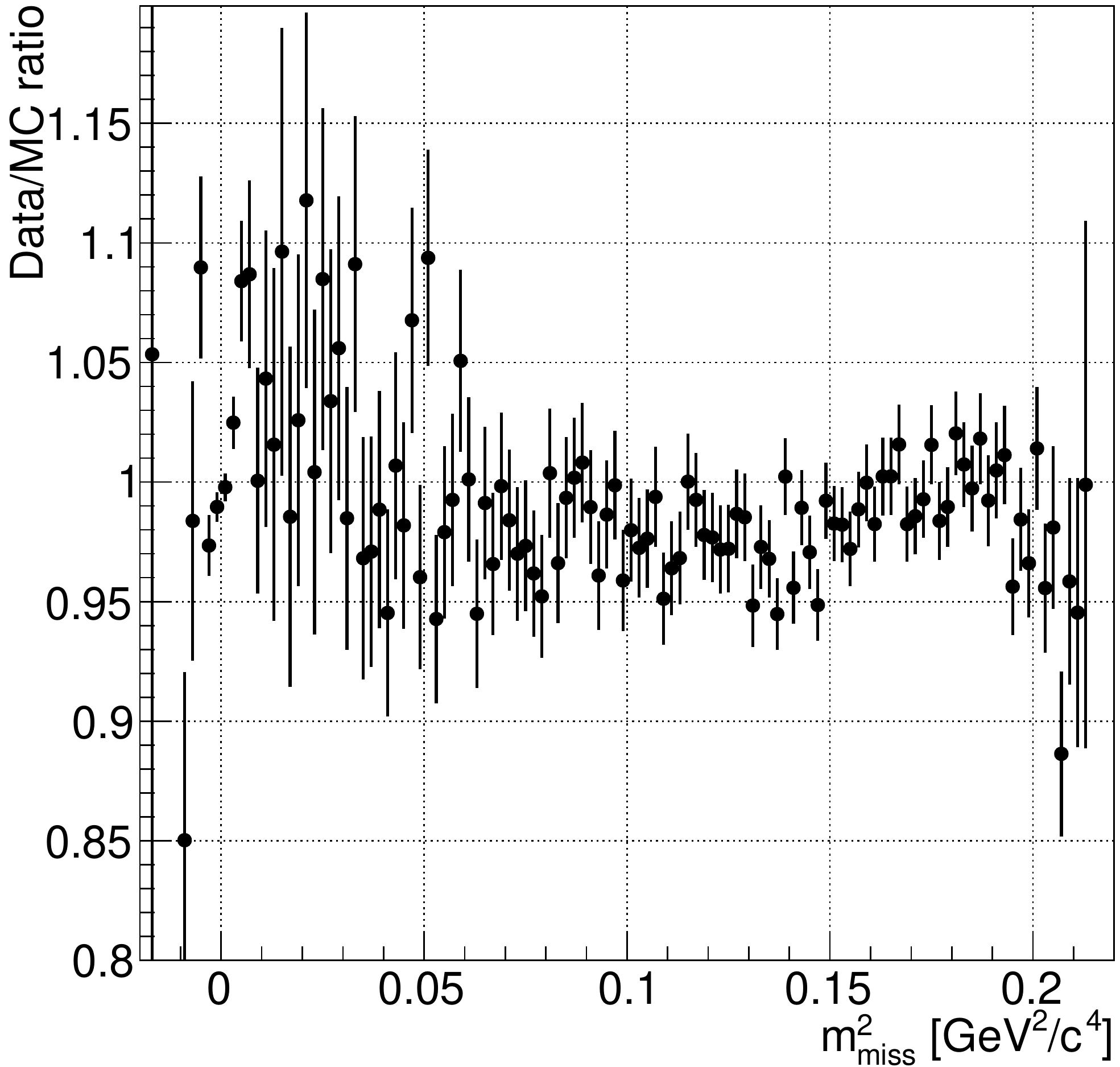}}
\end{center}
\vspace{-14mm}
\caption{Left: reconstructed squared missing mass ($m_{\rm miss}^2$) distributions for data and simulated events after the selection described in Section~\ref{sec:selection}. Decays to muons contribute via muon decays in flight. The boundaries of the SM signal region (upper arrows) and the HNL search region (lower arrows) are defined in Sections~\ref{sec:selection} and \ref{sec:search}. Right: the ratio of data and simulated $m_{\rm miss}^2$ spectra. The statistical uncertainties shown are dominated by those of the simulated spectra.}
\label{fig:mmiss2}
\end{figure}

Simulation of particle interactions with the detector and its response is performed with a Monte-Carlo simulation package based on the \geant toolkit~\cite{geant4}. The $m_{\rm miss}^2$ spectra of the events selected from data and simulated samples, and their ratio, are displayed in Fig.~\ref{fig:mmiss2}. The signal from the SM leptonic decay \mbox{$K^+\to e^+\nu$} is observed as a peak at $m_{\rm miss}^2=0$ with a resolution of \mbox{$1.7\times10^{-3}~{\rm GeV}^2/c^4$}. The simulation is tuned to reproduce this resolution to a 1\% relative precision. The SM signal region is defined in terms of the reconstructed squared missing mass as \mbox{$|m_{\rm miss}^2|<0.01~{\rm GeV}^2/c^4$}.

\section{Measurement principle}
\label{sec:flux}

A peak-search procedure measures the $K^+\to e^+N$ decay rate with respect to the
$K^+\to e^+\nu$ rate for an assumed HNL mass $m_N$. This approach benefits from first-order cancellations of residual detector inefficiencies not fully accounted for in simulations, as well as trigger inefficiencies and random veto losses, common to signal and normalization modes. The expected number of $K^+\to e^+N$ signal events $N_S$ can be written as
\begin{equation}
N_S = {\cal B}(K^+\to e^+N) / {\cal B}_{\rm SES}(K^+\to e^+N) = |U_{e4}|^2/|U_{e4}|^2_{\rm SES},
\label{eq:master}
\end{equation}
where the branching fraction ${\cal B}_{\rm SES}(K^+\to e^+N)$ and the mixing parameter $|U_{e4}|^2_{\rm SES}$ corresponding to the observation of one signal event, the single event sensitivity (SES), are defined as
\begin{equation}
{\cal B}_{\rm SES}(K^+\to e^+N) = \frac{1}{N_K \cdot A_N} ~~~~ {\rm and} ~~~~
|U_{e4}|^2_{\rm SES} = \frac{{\cal B}_{\rm SES}(K^+\to e^+N)}{{\cal B}(K^+\to e^+\nu) \cdot \rho_e (m_N)},
\label{eq:ses}
\end{equation}
where $N_K$ is the number of $K^+$ decays in the FV, $A_N$ is the signal selection acceptance, and the kinematic factor $\rho_e(m_N)$ is defined in Eq.~(\ref{eq:rho}).

The number of $K^+$ decays in the FV is evaluated using the number of $K^+\to e^+\nu$ candidates reconstructed in the data sample. Data losses due to trigger efficiencies and random vetoes are included in the $N_K$ definition, which makes the value of $N_K$ specific to this analysis. The dominant background due to $K^+\to\mu^+\nu$ decay followed by $\mu^+\to e^+\nu\bar\nu$ decay (0.08\% in relative terms) is taken into account. Other backgrounds, including the contribution from $K^+\to\mu^+\nu$ decay with a misidentified muon, are negligible.

The number of $K^+$ decays is computed as
\begin{displaymath}
N_K = \frac{N_{\rm SM}}{A_e \cdot {\cal B}(K^+\to e^+\nu) + A_\mu \cdot {\cal B}(K^+\to\mu^+\nu)} = (3.52\pm0.02)\times 10^{12},
\end{displaymath}
where $N_{\rm SM}=3.495\times 10^{6}$ is the number of selected data events in the SM signal region; $A_e=(6.27\pm 0.02_{\rm stat})\times 10^{-2}$ and $A_\mu=(1.24\pm0.19_{\rm stat})\times 10^{-9}$ are the acceptances of the selection for the $K^+\to e^+\nu$ decay and the $K^+\to\mu^+\nu$ decay (followed by muon decay) evaluated with simulations; and
${\cal B}(K^+\to e^+\nu)=(1.582\pm0.007)\times 10^{-5}$ and ${\cal B}(K^+\to\mu^+\nu)=0.6356\pm0.0011$ are the branching fractions of these decays~\cite{pdg}. The uncertainty quoted in $N_K$ is due to the precision of the external input ${\cal B}(K^+\to e^+\nu)$ and the statistical and systematic accuracy of the simulation. The systematic uncertainty is evaluated by varying the selection criteria.


\section{Background evaluation with simulations}
\label{sec:bkg}

The search procedure is based on a data-driven estimation of the background to $K^+\to e^+N$ decays, which is valid in the absence of peaking signal-like background structures in the reconstructed mass spectrum. Simulations are used to understand the background qualitatively, to optimize the event selection, and to justify the search procedure.
%
%

The main background to $K^+\to e^+N$ and $K^+\to e^+\nu$ decays comes from $K^+\to\mu^+\nu$ decay followed by muon decay $\mu^+\to e^+\nu\bar\nu$. This background is reduced by the requirement of spatial compatibility between the positron and kaon tracks, and its ultimate level is limited by the resolution on the track directions provided by the STRAW and GTK spectrometers. The background from $K^+\to\mu^+\nu$ decays upstream of the vacuum tank is 60~times smaller than that from decays in the vacuum tank.

Backgrounds from beam pion decays $\pi^+\to e^+\nu$ and $\pi^+\to\mu^+\nu$ decays,
followed by muon decay, arise from $\pi^+$ misidentification in the KTAG due to the presence of an in-time beam kaon not decaying in the FV. Considering the beam rate, the kaon fraction in the beam, and the KTAG--RICH timing conditions used, the beam pion misidentification probability due to pileup is about 6\%. The probability of beam pion identification as a kaon in the KTAG in the absence of pileup is negligible. The beam pion decay background populates the missing mass region $m_{\rm miss}^2>0.13~{\rm GeV}^2/c^4$, because of the 30~GeV/$c$ upper limit imposed on the $e^+$ momentum.

The reconstructed $m_{\rm miss}^2$ spectrum of data events is described by simulations to a few percent relative precision (Fig.~\ref{fig:mmiss2}). The accuracy of the simulation is limited by systematic effects, such as in the modelling of the LAV detector response to soft radiative photons.


\begin{figure}[t]
\begin{center}
\resizebox{0.50\textwidth}{!}{\includegraphics{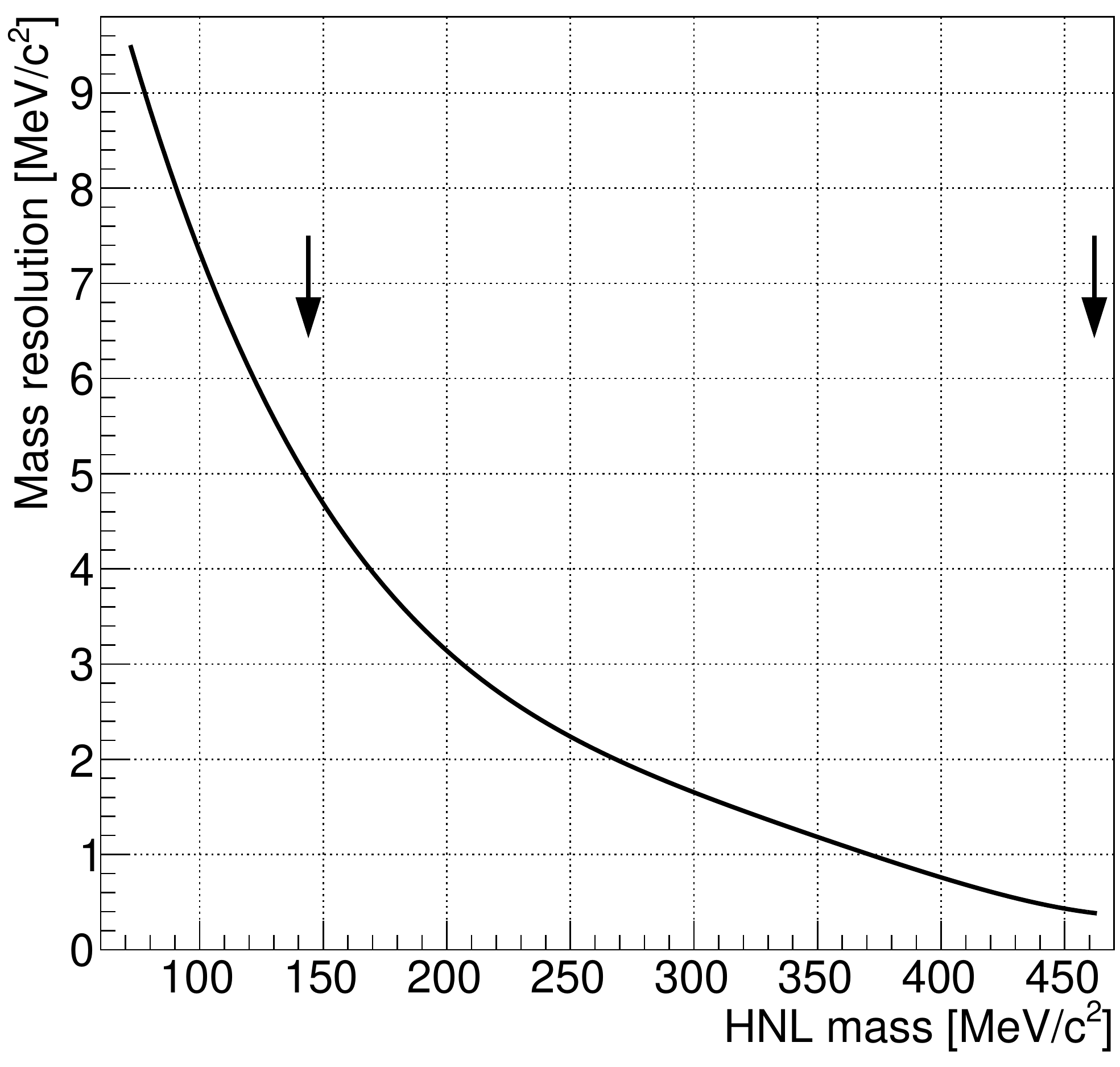}}%
\resizebox{0.50\textwidth}{!}{\includegraphics{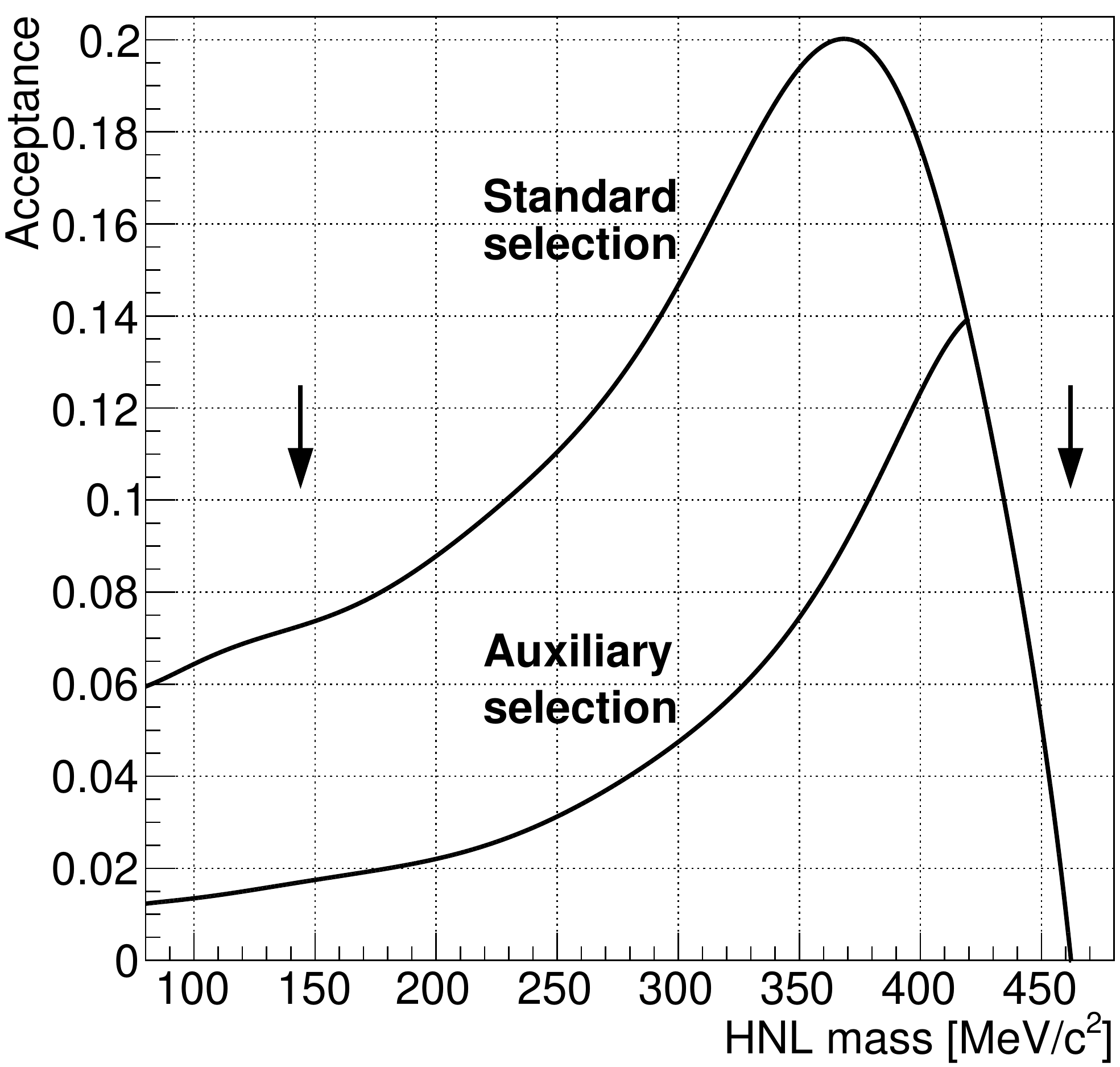}}
\end{center}
\vspace{-15mm}
\caption{Missing mass resolution $\sigma_m$ (left) and acceptances $A_N$ of the standard and auxiliary selections (right) evaluated from simulations as functions of the HNL mass, obtained by polynomial fits to measurements based on 80 simulated signal samples with different HNL masses. Above the mass of 420~MeV/$c^2$, the two selections have equal acceptance as the momentum of the positron produced in the $K^+\to e^+N$ decay is always below 20~GeV/$c$. The boundaries of the HNL search region are indicated by vertical arrows.}
\vspace{-1mm}
\label{fig:resolution-acceptance}
\end{figure}


\section{Search procedure}
\label{sec:search}

The $K^+\to e^+N$ process is investigated in 264 mass hypotheses, $m_N$, within the HNL search region between 144 and 462 MeV/c$^2$. The distances between adjacent mass hypotheses are equal to the mass resolution $\sigma_m$ shown in Fig.~\ref{fig:resolution-acceptance}~(left), rounded to 0.1~MeV/$c^2$. This mass resolution is three times better than that of the 2015 data sample collected without the GTK spectrometer~\cite{co18}. Event selection requires that $|m_{\rm miss}-m_N|<1.5\sigma_m$ for each mass hypothesis, where $m_{\rm miss}$ is the reconstructed missing mass.

In each HNL mass hypothesis, sidebands are defined in the reconstructed missing mass spectrum as $1.5\sigma_m<|m_{\rm miss}-m_N|<11.25\sigma_m$, additionally requiring the missing mass to be within the range 122--465~MeV/$c^2$. The number of expected background events, $N_{\rm exp}$, within the $\pm1.5\sigma_m$ signal window is evaluated with a second-order polynomial fit to the sideband data of the $m_{\rm miss}$ spectrum, where the bin size is $0.75\sigma_m$. The uncertainty, $\delta N_{\rm exp}$, in the number of expected background events includes statistical and systematic components. The former comes from the statistical errors in the fit parameters, and the latter is evaluated as the difference between $N_{\rm exp}$ obtained from fits using second and third order polynomials. The dominant contribution to $\delta N_{\rm exp}$ is statistical, except near the boundaries of the HNL search region where the systematic uncertainty is comparable. Further systematic errors due to possible HNL signals in the sidebands are found to be negligible; this check is made assuming $|U_{e4}|^2$ to be equal to the expected sensitivity of the analysis. The ratio $\delta N_{\rm exp}/N_{\rm exp}$ is typically 0.2--0.3\%, but reaches a few percent close to the limits of the search region.

An auxiliary selection with a tighter maximum positron momentum requirement of 20~GeV/$c$ is used to achieve a locally smooth $m_{\rm miss}$ spectrum of background events in the sidebands for mass hypotheses in the range 356--382~MeV/$c^2$. The beam pion decay background threshold in the $m_{\rm miss}^2$ spectrum is shifted from $0.14~{\rm GeV}^2/c^4$ (Fig.~\ref{fig:mmiss2}) to $0.165~{\rm GeV}^2/c^4$ within the auxiliary selection because the $e^+$ momentum and the $m_{\rm miss}^2$ reconstructed for the $\pi^+\to e^+\nu$ decay in the $K^+$ mass hypothesis are anti-correlated.


The signal selection acceptances, $A_N$, for the standard and the auxiliary selections as functions of $m_N$ obtained with simulations assuming infinite HNL lifetime are displayed in Fig.~\ref{fig:resolution-acceptance}~(right). The acceptance for a mean lifetime of 50~ns (considering decays to detectable particles) is lower than shown by ${\cal O}(1\%)$ in relative terms, making the results of the search valid for lifetimes in excess of 50~ns. For smaller lifetimes, the HNL mean decay length in the laboratory frame becomes comparable to or smaller than the length of the apparatus. Consequently, acceptances for lifetimes of 5~(1)~ns decrease due to the veto conditions by factors of up to 2~(10), depending on the HNL mass.

The simulation of the missing mass resolution outside the $K^+\to e^+\nu$ peak is validated with a sample of fully reconstructed $K^+\to\pi^+\pi^+\pi^-$ decays by studying the resolution of $\Delta m_{3\pi} = \sqrt{(P_K-P_3)^2}-\sqrt{(P_1+P_2)^2}$, where $P_{i}$ ($i=1,2,3$) and $P_K$ are the reconstructed pion and kaon 4-momenta. The resolution on $\Delta m_{3\pi}$ can be measured for both data and simulated samples because the true value of this quantity is always zero. The resolution varies in the range 0.8--1.3~MeV/$c^2$ depending on $(P_1+P_2)^2$, and the simulation agrees with the data to better than 5\%, validating the signal acceptance estimates to 2\% relative precision.

\begin{figure}[t]
\begin{center}
\resizebox{0.50\textwidth}{!}{\includegraphics{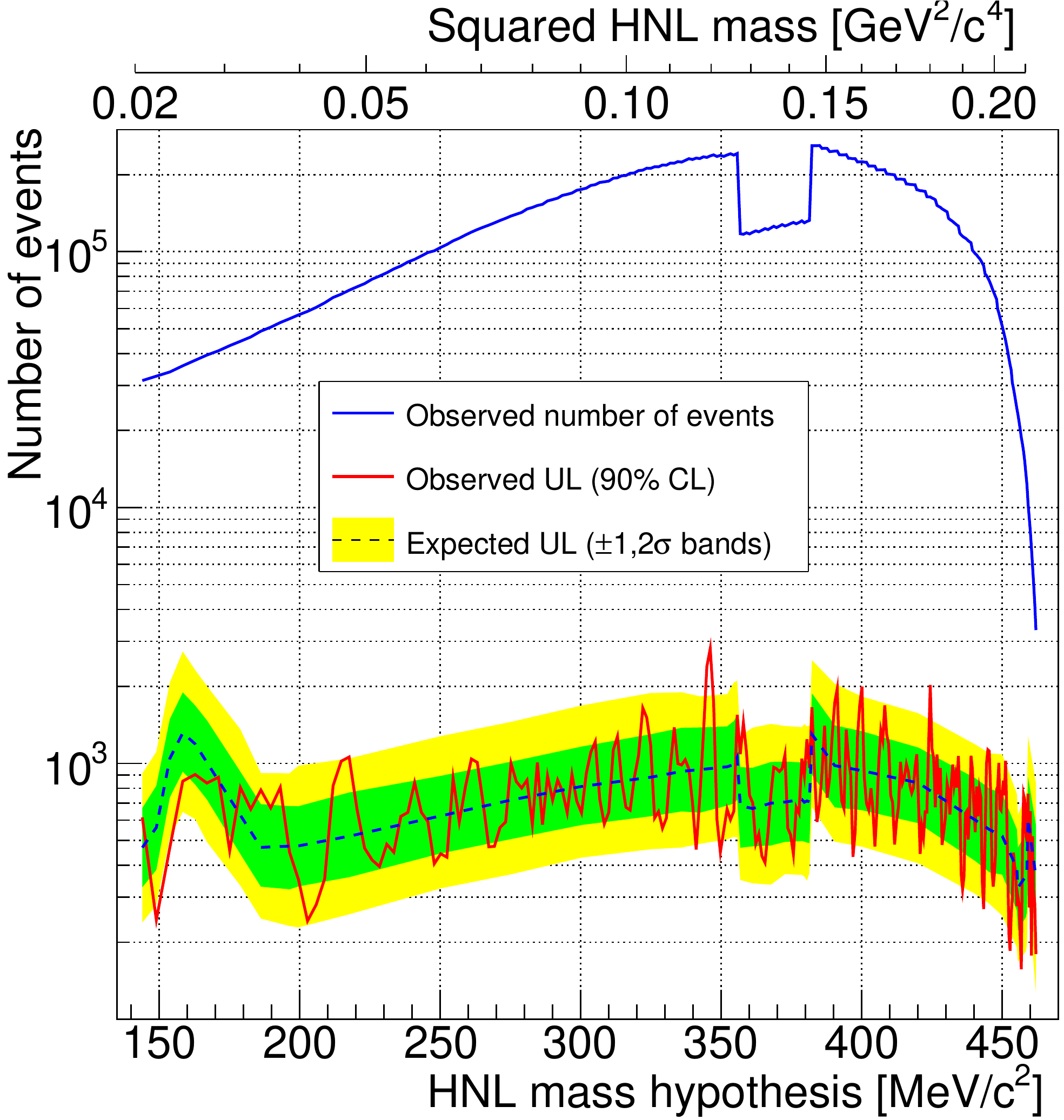}}%
\resizebox{0.50\textwidth}{!}{\includegraphics{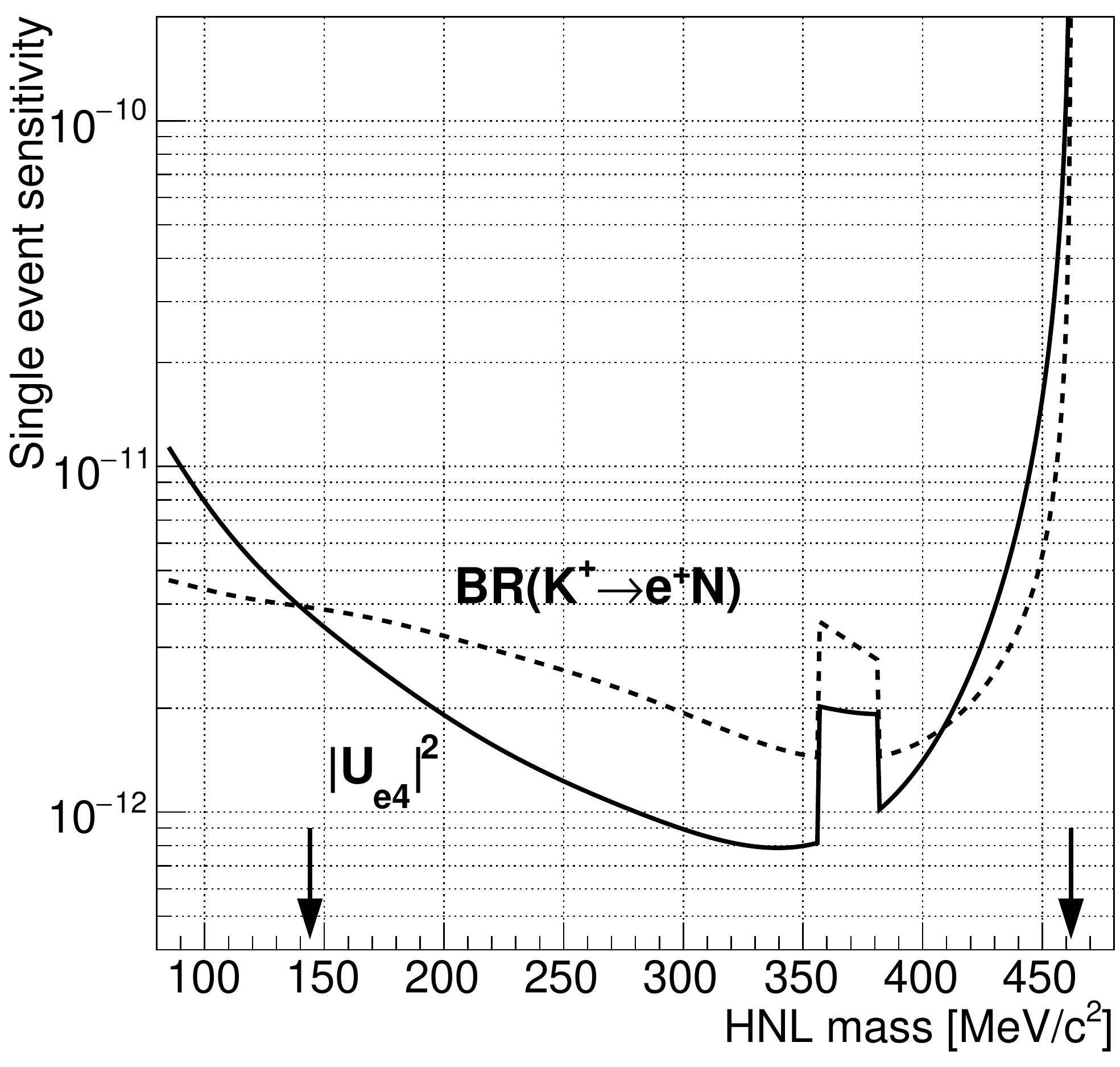}}%
\boldmath
\unboldmath
\end{center}
\vspace{-13mm}
\caption{Left: the values of $N_{\rm obs}$, the obtained upper limits at 90\% CL of the numbers of $K^+\to e^+N$ events, and the expected $\pm1\sigma$ and $\pm2\sigma$ bands in the background-only hypothesis for each HNL mass value considered. Right: single event sensitivities ${\cal B}_{\rm SES}(K^+\to e^+N)$ (dashed line) and $|U_{e4}|^2_{\rm SES}$ (solid line) as functions of the HNL mass. Note the reduced sensitivity in the region of 356--382~MeV/$c^2$ in which the auxiliary selection is used. The boundaries of the HNL search region are indicated by vertical arrows.}
\label{fig:ses}
\end{figure}


\section{Results}
\label{sec:results}

The number of observed events, $N_{\rm obs}$, within the signal window, the number of expected background events, $N_{\rm exp}$, and its uncertainty, $\delta N_{\rm exp}$, are used to compute the local signal significance for each mass hypothesis
\begin{displaymath}
z=(N_{\rm obs}-N_{\rm exp})/\sqrt{(\delta N_{\rm obs})^2+(\delta N_{\rm exp})^2},
\end{displaymath}
with $\delta N_{\rm obs}=\sqrt{N_{\rm obs}}$. A maximum local significance of 3.6 is found for $m_N=346.1~{\rm MeV}/c^2$, based on $N_{\rm obs}=236745$ and $N_{\rm exp}=234678\pm314$. Accounting for the look-elsewhere effect, the global significance becomes 2.2.

The quantities $N_{\rm obs}$, $N_{\rm exp}$, and $\delta N_{\rm exp}$ are used to evaluate the upper limit at 90\% CL of the number of $K^+\to e^+N$ decays, $N_S$, in each HNL mass hypothesis using the ${\rm CL_S}$ method~\cite{re02}. The values of $N_{\rm obs}$, the obtained upper limits of $N_S$, and the expected $\pm1\sigma$ and $\pm 2\sigma$ bands of variation of $N_S$ in the background-only hypothesis are shown in Fig.~\ref{fig:ses}~(left).

\begin{figure}[t]
\begin{center}
\resizebox{0.50\textwidth}{!}{\includegraphics{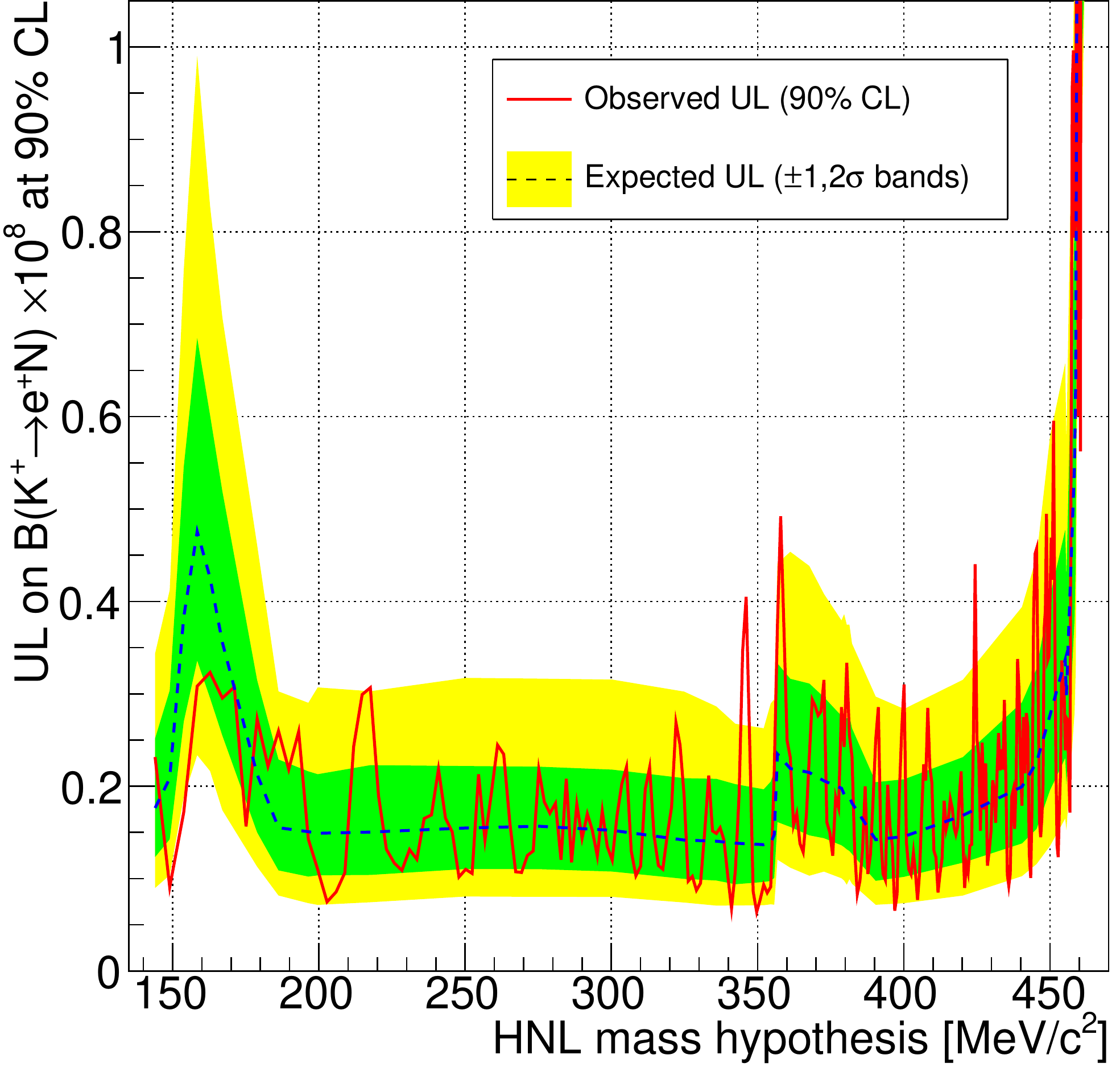}}
\end{center}
\vspace{-13mm}
\caption{The obtained upper limits at 90\% CL of ${\cal B}(K^+\to e^+N)$ and the expected $\pm1\sigma$ and $\pm2\sigma$ bands in the background-only hypothesis for each HNL mass value considered.}
\label{fig:br}
\end{figure}

\begin{figure}[t]
\begin{center}
\resizebox{0.7\textwidth}{!}{\includegraphics{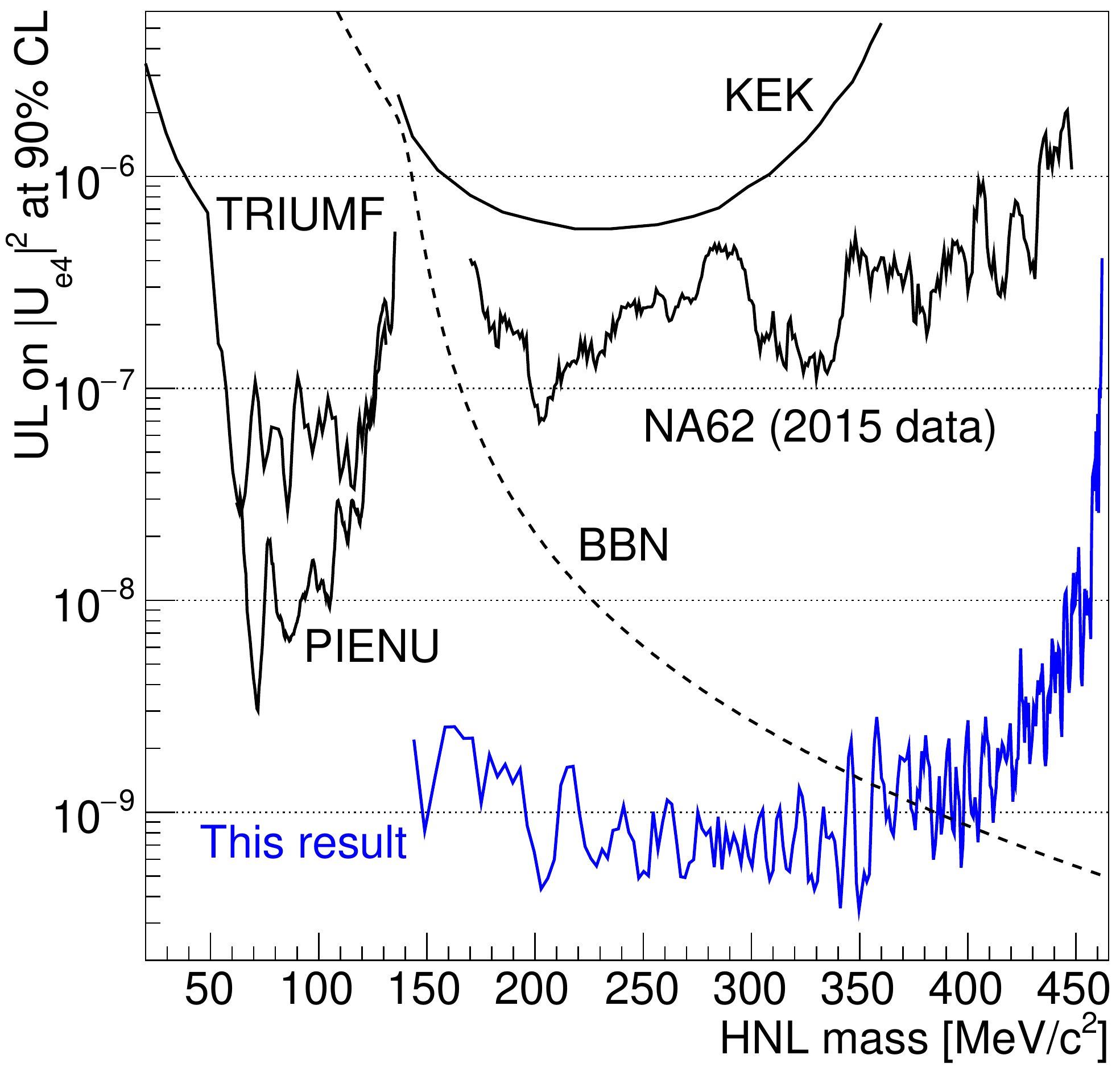}}
\end{center}
\vspace{-13mm}
\caption{Upper limits at 90\% CL of $|U_{e4}|^2$ obtained for each assumed HNL mass compared to the limits established by earlier HNL production searches in $K^+\to e^+N$ decays: KEK~\cite{ya84}, NA62 (2015 data)~\cite{co18}; and $\pi^+\to e^+N$ decays: TRIUMF~\cite{br92}, PIENU~\cite{ag18}. The lower boundary on $|U_{e4}|^2$ imposed by the BBN constraint~\cite{do00} is shown by a dashed line.}
\label{fig:world}
\end{figure}

Upper limits at 90\% CL of the branching fraction ${\cal B}(K^+\to e^+N)$ and the mixing parameter $|U_{e4}|^2$ are obtained from those of $N_S$ according to Eq.~(\ref{eq:master}), using the single event sensitivities ${\cal B}_{\rm SES}(K^+\to e^+N)$ and $|U_{e4}|^2_{\rm SES}$ shown as functions of the HNL mass in Fig.~\ref{fig:ses} (right). The obtained limits of ${\cal B}(K^+\to e^+N)$ are displayed in Fig.~\ref{fig:br}. The obtained limits of $|U_{e4}|^2$, together with the limits from previous HNL production searches in $K^+$~\cite{co18,ya84} and $\pi^+$~\cite{br92,ag18} decays in the 30--470~MeV/$c^2$ mass range, and the Big Bang Nucleosynthesis (BBN) constraint~\cite{do00}, are shown in Fig.~\ref{fig:world}. The reported result improves the existing limits of $|U_{e4}|^2$ obtained in production searches over the whole mass range considered. Comparison with HNL decay searches is available in Refs.~\cite{pbc19,bo19}. The obtained limits of $|U_{e4}|^2$ improve over the decay searches~\cite{ps191,abe19} in the mass region below 400~MeV/$c^2$.

\section*{Summary and outlook}

A search for HNL production in $K^+\to e^+N$ decays has been performed with the data set collected by the NA62 experiment at CERN in 2017--2018. Upper limits of the decay branching fraction and the mixing parameter $|U_{e4}|^2$ have been established at the $10^{-9}$ level over most of the HNL mass range 144--462~MeV/$c^2$ with the assumption of mean lifetime exceeding 50~ns. These limits become weaker by factors of up to 2~(10) for lifetimes of 5~(1)~ns. The $|U_{e4}|^2$ results are significantly better than previous limits obtained from HNL production and decay searches~\cite{pbc19}, and other experimental constraints~\cite{bo19,br19}. The values of $|U_{e4}|^2$ favoured by the BBN constraint~\cite{do00} are excluded for HNL masses up to about 340~MeV/$c^2$.

An improvement in sensitivity of this analysis in terms of $|U_{e4}|^2$ can only be expected with future NA62 data. Considering the background conditions, the sensitivity to $|U_{e4}|^2$ is proportional to $\delta N_{\rm exp}/N_{\rm exp}$, and improves as the relative statistical uncertainty in $N_{\rm exp}$ decreases with sample size as $1/\sqrt{N_K}$.

HNL masses below 144~MeV/$c^2$, not accessible to this analysis due to the shape of the background mass spectrum, can be probed via the $K^+\to\pi^0 e^+N$ decay. The data set currently available for this search, collected with a pre-scaled trigger, corresponds to $N_K\approx 3\times 10^{10}$. The single event sensitivity in the mass range of interest estimated using the formalism of Ref.~\cite{bo18} is $|U_{e4}|_{\rm SES}^2\approx 10^{-8}$; the search is expected to be limited by the $K^+\to\pi^0e^+\nu\gamma$ background.


\section*{Acknowledgements}

It is a pleasure to express our appreciation to the staff of the CERN laboratory and the technical staff of the participating laboratories and universities for their efforts in the operation of the experiment and data processing. We are grateful to Matheus Hostert, Silvia Pascoli, Oleg Ruchayskiy, Robert Shrock, Jean-Loup Tastet and Inar Timiryasov for the inputs provided on the HNL phenomenology.

The cost of the experiment and its auxiliary systems was supported by the funding agencies of
the Collaboration Institutes. We are particularly indebted to:
F.R.S.-FNRS (Fonds de la Recherche Scientifique - FNRS), Belgium;
BMES (Ministry of Education, Youth and Science), Bulgaria;
NSERC (Natural Sciences and Engineering Research Council), funding SAPPJ-2018-0017 Canada;
NRC (National Research Council) contribution to TRIUMF, Canada;
MEYS (Ministry of Education, Youth and Sports),  Czech Republic;
BMBF (Bundesministerium f\"{u}r Bildung und Forschung) contracts 05H12UM5, 05H15UMCNA and 05H18UMCNA, Germany;
INFN  (Istituto Nazionale di Fisica Nucleare),  Italy;
MIUR (Ministero dell'Istruzione, dell'Universit\`a e della Ricerca),  Italy;
CONACyT  (Consejo Nacional de Ciencia y Tecnolog\'{i}a),  Mexico;
IFA (Institute of Atomic Physics) Romanian CERN-RO No.1/16.03.2016 and Nucleus Programme PN 19 06 01 04,  Romania;
INR-RAS (Institute for Nuclear Research of the Russian Academy of Sciences), Moscow, Russia;
JINR (Joint Institute for Nuclear Research), Dubna, Russia;
NRC (National Research Center)  ``Kurchatov Institute'' and MESRF (Ministry of Education and Science of the Russian Federation), Russia;
MESRS  (Ministry of Education, Science, Research and Sport), Slovakia;
CERN (European Organization for Nuclear Research), Switzerland;
STFC (Science and Technology Facilities Council), United Kingdom;
NSF (National Science Foundation) Award Numbers 1506088 and 1806430,  U.S.A.;
ERC (European Research Council)  ``UniversaLepto'' advanced grant 268062, ``KaonLepton'' starting grant 336581, Europe.

Individuals have received support from:
Charles University Research Center (UNCE/SCI/013), Czech Republic;
Ministry of Education, Universities and Research (MIUR  ``Futuro in ricerca 2012''  grant RBFR12JF2Z, Project GAP), Italy;
Russian Foundation for Basic Research  (RFBR grants 18-32-00072, 18-32-00245), Russia;
Russian Science Foundation (RSF 19-72-10096), Russia;
the Royal Society  (grants UF100308, UF0758946), United Kingdom;
STFC (Rutherford fellowships ST/J00412X/1, ST/M005798/1), United Kingdom;
ERC (grants 268062,  336581 and  starting grant 802836 ``AxScale'');
EU Horizon 2020 (Marie Sk\l{}odowska-Curie grants 701386, 842407, 893101).




\end{document}